\documentclass[preprint]{revtex4}

\usepackage{ifpdf}

\ifpdf{}
\usepackage[pdftex]{graphicx}
\usepackage[pdftex,colorlinks,linkcolor=black]{hyperref}
\else{}
\usepackage{graphicx}
\usepackage[colorlinks,linkcolor=black]{hyperref}
\fi{}

\usepackage{amsmath,units,textcomp}
\usepackage{ragged2e}

\graphicspath{{./}{figs/}{figs/output/}}

\newcommand{\lee}[1]{\ensuremath{\lambda_{e-e}^{#1}}}
\newcommand{\leem}[1]{\ensuremath{\left\langle\lambda_{e-e}^{#1}\right\rangle}}

\begin{document}


\title{Direct Measurement of 2D and 3D Interprecipitate Distance Distributions
from Atom-Probe Tomographic Reconstructions}

\author{Richard A. Karnesky}
\email{karnesky@northwestern.edu}
\author{Dieter Isheim}
\author{David N. Seidman}
\affiliation{
Department of Materials Science and Engineering and the Northwestern University Center for Atom-Probe Tomography (NUCAPT)\\
Evanston, IL 60208--3108, USA}
\homepage{http://arc.nucapt.northwestern.edu/}

\date{2007-04-30}

\begin{abstract}
Edge-to-edge interprecipitate distance distributions are critical for predicting
precipitation strengthening of alloys and other physical phenomena.  A method to
calculate this 3D distance and the 2D interplanar distance from atom-probe
tomographic data is presented.  It is applied to nanometer-sized Cu-rich
precipitates in an Fe-1.7~at.\%~Cu alloy.  Experimental interprecipitate
distance distributions are discussed.
\end{abstract}

\maketitle
 
Many physical properties of materials depend on the edge-to-edge
interprecipitate distance, \lee{}.  The applied stress required for a
dislocation to glide past or climb over precipitates depends on
\lee{}~\cite{Brown-1971}, as does precipitate coarsening and electrical
conductivity.  Frequently, \lee{} is merely approximated by assuming the
precipitates form a cubic array or a square array in a
plane~\cite{Nembach-1996}.  It is also assumed that precipitates are spherical
with a known precipitate size distribution (PSD) (usually either all
precipitates are the same size or they obey the PSD derived by Lifshitz and
Slyozov~\cite{Lifshitz-1961} and Wagner~\cite{Wagner-1961} (LSW)).  Real
materials are almost always more complicated.

Much of the past work on calculating the distance between precipitates or other microstructural features of interest~\cite{Torquato-2002,Quintanilla-2004}%
\ (whether interprecipitate distances~\cite{Nembach-1996}, mean free paths or
chord lengths~\cite{Lu-1993}, or nearest-neighbor distribution
functions~\cite{Torquato-1995,Macdonald-1996,Liu-2001,Tewari-2004}) has been
theoretical.  Experimental characterization of \lee{} requires a microscopic
technique that has: (i) a high enough spatial resolution to define clearly each
and every precipitate; (ii) a large enough analysis volume to capture many
precipitates and to exclude boundary effects; and (iii) 3-dimensional
information (without suffering from precipitate overlap or truncation).  For
nanometer-sized precipitates, the local-electrode atom-probe%
\ (LEAP\textsuperscript{\textregistered}) tomograph (Imago Scientific
Instruments) satisfies these requirements~\cite{Kelly-2000,Seidman-2007}.
Despite these capabilities, it has not been previously utilized to gather this
information and the little available experimental data for \lee{} comes from 2D
techniques.  These cannot be compared directly to models of 3D microstructure,
but only to 2D slices from theoretical 3D microstructures~\cite{Leggoe-2005}.

In this article, an algorithm to calculate \lee{} from LEAP tomographic
reconstructions is presented and applied to a binary Fe-Cu alloy.  This alloy
and many other steels are strengthened by a high number density of
nanometer-sized copper-rich precipitates~\cite{Lahiri-1970}.  Many of the
proposed precipitate strengthening mechanisms depend on \lee{}~\cite{Fine-2005,%
Liu-2005}.  Atom-probe tomography has been used to study the size, morphology,
and chemical composition of Cu precipitates~\cite{Goodman-1973,Miller-1998,%
Murayama-1999,Isheim-2006,Isheim-2006_2}, but not to measure \lee{}.
\begin{figure}[Htb]
  \begin{center}
    \includegraphics[width=8.5cm]{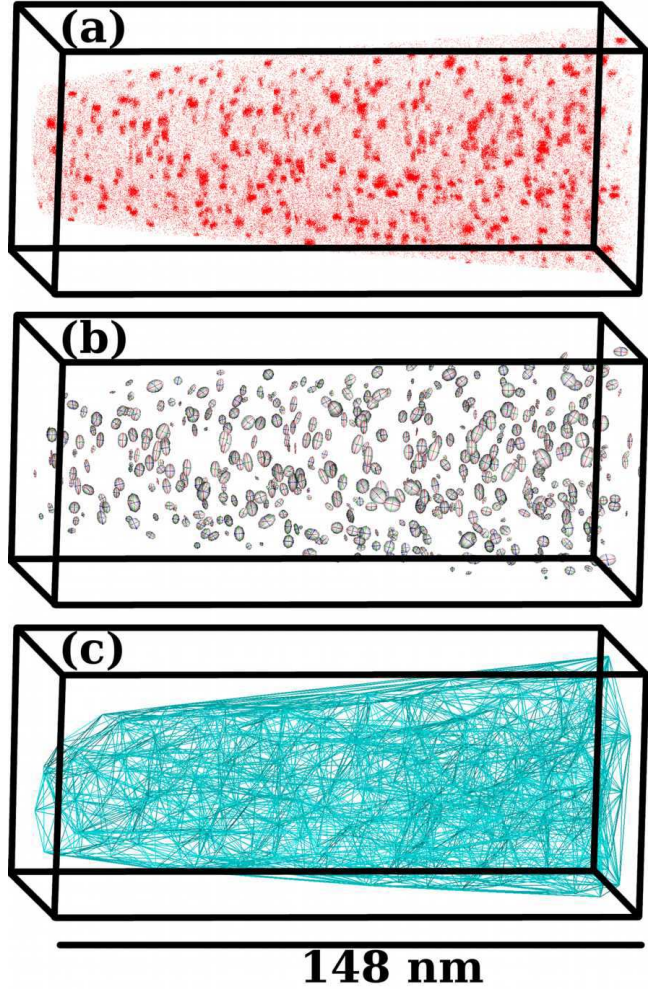}
  \end{center}
  \caption{(a) A LEAP\textsuperscript{\textregistered} tomographic
  reconstruction of an Fe-1.7~at.\%~Cu specimen, whose thermal history is
  detailed in the text.  Only Cu atoms are displayed for clarity.
  (b) The 546 precipitates are fitted as ellipsoids~\cite{Karnesky-2007}
  (c) A Delaunay mesh connects the precipitate centers.  This is used to find
  ``interacting precipitates'' and to exclude the convex hull.}\label{fig:apt}
\end{figure}
An Fe-1.7~at.\%~Cu alloy was solutionized at 1000\textcelsius{} for 1 h and
845\textcelsius{} for 6 h.  It was subsequently aged for 2 h at
500\textcelsius{}.  This treatment leads to a high number density%
\ ($\mathrm{(1.2\pm0.1)\times10^{24}~m^{-3}}$) of nanometer-sized precipitates%
\ (with a mean radius, $\langle{R}\rangle$, equal to $\mathrm{1.3~nm}$).  The
specimens were cut, ground, and then electropolished into tips.  The LEAP
tomographic experiment was conducted with a 50~K specimen temperature, a
5--10~kV specimen voltage, pulse fraction of 15\%, and a pulse repetition rate
of 200~kHz to collect ca.~$\mathrm{1.3\times10^6}$~ions in a
$\mathrm{148\times66\times62~nm^3}$~volume (Fig.~\ref{fig:apt}a).  The computer
program \textsc{ivas} (Imago Scientific Instruments) was used to analyze the
data.  Precipitates are isolated using a modified envelope
algorithm~\cite{Miller-2004}.  Because Cu partitions strongly to
precipitates~\cite{Isheim-2006}, an isoconcentration surface was not necessary
to distinguish the 546 precipitates in this dataset.

The interprecipitate distance algorithm begins by representing these
precipitates with simpler geometric shapes.  While \lee{} between spheres is
simple (it being the difference of the center-to-center distance and the
precipitate radii), spheres do not adequately represent many precipitate
morphologies.  Instead, best-fit ellipsoids to the precipitates are calculated%
\ (Fig.~\ref{fig:apt}b) employing a recently presented
algorithm~\cite{Karnesky-2007}.  The $\mathrm{4\times4}$ transformation matrix
calculated with that algorithm translates, rotates, and scales a unit sphere
centered at the origin to an ellipsoid that preserves the centroid, principle
axes, and moments of inertia of a precipitate.

A Delaunay tedrahedral mesh is generated from the precipitate centroids%
\ (Fig.~\ref{fig:apt}c)~\cite{Clarkson-1992,Barber-1996}.  The Delaunay mesh is
the geometric dual of the Voronoi diagram; mesh segments connect neighboring
precipitates whose Voronoi cells touch.  It decreases the number of precipitate
pairs for which \lee{} is calculated to a group of neighbors.  The mesh also
finds the 75 precipitates that make up the convex hull.  These outer-most
precipitates are allowed to be nearest neighbors of the inner precipitates, but
their own nearest neighbors are not calculated, as they might fall outside the
volume of the analysis.

The distance between two ellipsoids is found utilizing the constrained
optimization by linear approximation (COBYLA) algorithm~\cite{Powell-1994}. This
general optimization algorithm is chosen over more efficient algorithms that
calculate explicitly the distance between ellipsoids~\cite{Lin-2002,Sohn-2002},
so that it can be used with other abstractions of precipitate morphology (such
as the isoconcentration surface) and additional constraints (such as calculation
of interplanar edge-to-edge distances) and because a gratis implementation
exists~\cite{Jones-2001}.  COBYLA minimizes the distance between two points, $x$
and $y$ in the analysis space, $\sqrt{\sum_{j}\left(x_j-y_j\right)^2}$.  The
constraints are that $x$ and $y$ must fall on the ellipsoid.  This is simplified
by the fact that applying the inverse transformation of ellipsoids transforms
them back into unit spheres, centered at the origin (so $\sum_{j}x_j^{T_x}=1$
and $\sum_{j}y_j^{T_y}=1$, where the superscript $T_i$ is the inverse transform
of the best-fit ellipsoid for precipitate $i$).  The initial guess is chosen as
the two closest points that satisfy these constraints that lie on the line that
connects the precipitate centers.
\begin{figure}[Htb]
  \begin{center}
    \includegraphics{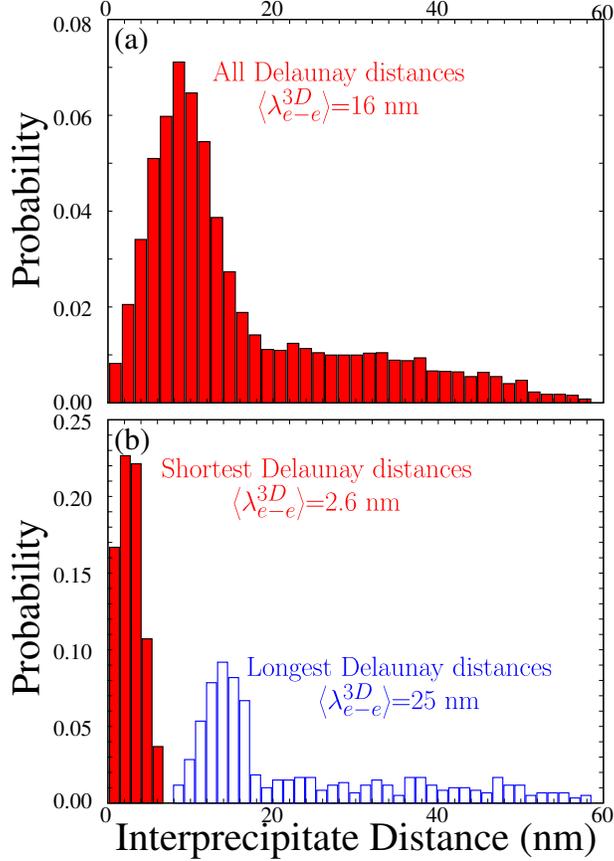}
  \end{center}
  \caption{3D IDD for the dataset in Fig.~\ref{fig:apt}.
  (a) IDD of all 6,771 Delaunay lengths, with \leem{3D}=16~nm.
  (b) Solid: IDD of nearest-neighbor distances, which is much sharper than when 
  longer lengths are included (\leem{3D}=2.6~nm).  Hollow: IDD of the
  most-distant Delaunay neighbors, which is broader than and does not overlap
  with the shortest distances (\leem{3D}=25~nm).}\label{fig:delaunay}
\end{figure}
Interprecipitate distance distributions (IDDs) may be generated using different
combinations of Delaunay neighbors, as in Fig.~\ref{fig:delaunay}.  An IDD is
the convolution of a PSD and the center-to-center distances.  In
Fig.~\ref{fig:delaunay}a, an IDD for all 6,671 Delaunay neighbor distances
yields a mean 3D interprecipitate distance, \leem{3D}, of 16~nm.
Figure~\ref{fig:delaunay}b displays two subsets of this IDD, each with 471
lengths.  The distance between nearest precipitates is often used to calculate
precipitate-dislocation interactions.  The IDD for this is much sharper and
\leem{3D}=2.6~nm.  Precipitates that are very close to one another might be
bypassed as a pair by a dislocation.  The longest Delaunay distances provide an
upper bound to the interactive distance.  This is probably not physically
important for plastic deformation, but may be relevant for other physical
phenomena.  The IDD for this case is broader, does not overlap the shortest
distances, and has a mean value that is an order of magnitude larger%
\ (\leem{3D}=25~nm).

In certain cases, it is not \leem{3D} that is of interest, but rather the
interplanar edge-to-edge distance, \leem{2D}.  This might, for instance, be a
glide plane of a dislocation.  This 2D distance can be calculated by imposing an
additional constraint for COBYLA---that $x$ and $y$ values must fall on a
particular plane.  For comparison, \leem{3D} and \leem{2D} can be calculated
from one another by assuming precipitates are distributed on a cubic
lattice~\cite{Nembach-1996},
\begin{equation}
  \leem{3D} = \left(\sqrt[3]{\frac{4}{3}\frac{\pi}{\phi}}-2\right)%
                 \langle{R}\rangle;\label{eq:3D}
\end{equation}
where $\phi$ is the volume fraction of precipitates.  Assuming a square array of
precipitates,
\begin{equation}
  \leem{2D} = \left(\sqrt{\frac{\pi}{\phi}}-2\right)\bar{R};\label{eq:2D}
\end{equation}
where the mean planar radius, $\bar{R}$, is equal to
$\frac{\pi}{4}\omega_2\langle{R}\rangle$, with $\omega_2$ dependent on the
PSD~\cite{Nembach-1996}.  Values for $\omega_2$ for the LSW distribution and for
the case where all precipitates are the same size are given in
Ref.~\cite{Nembach-1996}.  Equating the $\phi$s in Eqs.~\ref{eq:3D}--\ref{eq:2D}
leads to a cubic equation relating \leem{2D} and \leem{3D}.  Solving for
\leem{2D}:
\begin{equation}
  \leem{2D} = \frac{\pi\left(-4\omega_2\langle{R}\rangle^2+%
                   \sqrt{3\omega_2^2\langle{R}\rangle\left(2\langle{R}\rangle%
                    +\leem{3D}\right)^3}\right)}%
                    {8\langle{R}\rangle}.\label{eq:convert}
\end{equation}
\begin{figure}[Htb]
  \begin{center}
    \includegraphics{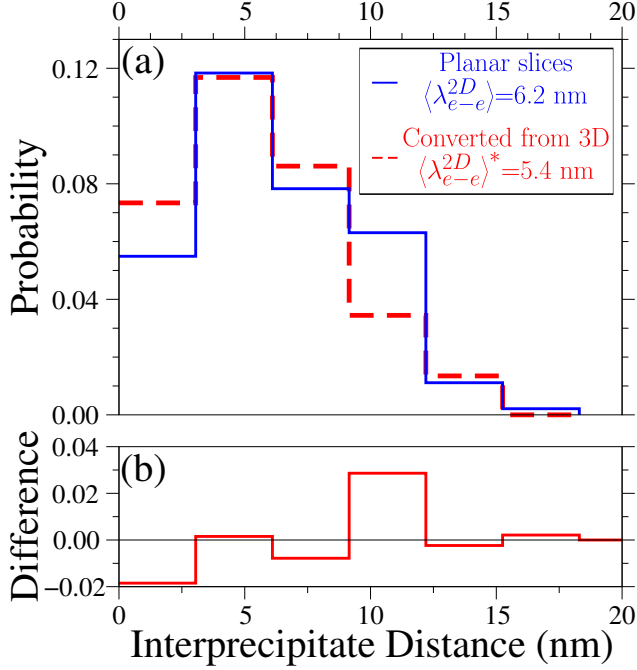}
  \end{center}
  \caption{(a) Interplanar (2D) IDDs for the dataset in Fig.~\ref{fig:apt}.
  Solid: IDD of slices, with \leem{2D}=6.2~nm.
  Dashed: IDD from Fig.~\ref{fig:delaunay}b scaled by Eq.~\ref{eq:convert}
  ($\leem{2D}^*$=5.4~nm).
  (b) \lee{2D} is weighted towards longer distances than
  ${\lee{2D}}^*$}\label{fig:convert}
\end{figure}
In Fig.~\ref{fig:convert}, the results of the two methods for extracting 2D
nearest-neighbor IDDs from the 3D dataset are compared.  \lee{2D} is calculated
directly by imposing the additional constraint on COBYLA that the two points
must lie in the same plane, which is radial to the analysis direction.  The
entire tip is sampled by taking 180 1\textdegree{} steps.  This process samples
the precipitates toward the center of the tip more than those toward the hull,
but has fewer edge artifacts than parallel slices would have.  ${\lee{2D}}^*$ is
calculated from what is displayed in Fig.~\ref{fig:delaunay}b by applying
Eq.~\ref{eq:convert} with $\langle{R}\rangle=\mathrm{1.3~nm}$ and
$\omega_2=1.046$~\cite{Nembach-1996} (the superscript $*$ denotes this
transformation).  Despite the simple geometrical assumptions involved in
deriving Eq.~\ref{eq:convert}, the mean values are in reasonable agreement (with
the ``direct'' method yielding \leem{2D}=6.2~nm and with the conversion leading
to $\leem{2D}^*$=5.4~nm).  Despite this similarity in the mean values, the
distributions are different. The converted IDD is narrower and weighted towards
shorter distances than the IDD that is directly calculated.

This supports the conclusion reached in Ref.~\cite{Leggoe-2005}, that $\lee{}$
should be calculated directly with the same dimensionality as either the
simulations they are compared with or the physical models they are be used in.
2D techniques can \emph{only} result in 2D IDDs, and will not yield accurate 3D
IDDs.  They may give a reasonable estimate of \leem{3D}, although 3D
experimental data, as is gathered with the LEAP tomograph, allows both \lee{3D}
to be measured and \lee{2D} to be measured from planar slices taken from the 3D
reconstruction.

We are in the midst of applying this approach to calculate the strength of
different alloys using analytical equations~\cite{Nembach-1996}.  We are also
using it to evaluate the statistical accuracy of simulated microstructures,
which are used in a continuum dislocation dynamics simulation that calculates a
stress-strain curve~\cite{Mohles-2004}.
\begin{acknowledgments}
This research is supported by the Office of Naval Research, under contract
N00014--03--1--0252.  RAK received partial support from a Walter P. Murphy
Fellowship and the US Department of Energy (DE--FG02--98ER45721).  We thank
Dr.\ S.\ K.\ Lahiri and Prof.\ M.\ E.\ Fine for providing the Fe-Cu alloy.
Imago Scientific Instruments and Dr.\ M.\ K.\ Miller permitted RAK to modify the
source code for \textsc{envelope}.  Profs.\ D.\ C.\ Dunand and J. Jerome are
thanked for discussions.
\end{acknowledgments}
{\RaggedRight{}

\end{document}